\documentclass[times,referee,usenatbib]{mn2e}
\usepackage{xspace}
\usepackage{psfig}
\usepackage{defs}
\newcommand{\be}{\begin{equation}}
\newcommand{\ee}{\end{equation}}
\newcommand{\bear}{\begin{eqnarray}}
\newcommand{\ear}{\end{eqnarray}}

\newcommand{\fig}{Figure}

\def\nCIV{{\rm CIV}}

\def\etal{{\it et al.~\/}}

\def\ie{{\it i.e.}}

\def\ltsima{$\; \buildrel < \over \sim \;$}
\def\simlt{\lower.5ex\hbox{\ltsima}}
\def\gtsima{$\; \buildrel > \over \sim \;$}
\def\simgt{\lower.5ex\hbox{\gtsima}}

\begin{document}

\title[Winds and Infall in LBGs]{Winds and Infalling Gas in Lyman Break Galaxies}
\author[A. Ferrara \& M. Ricotti]
{Andrea Ferrara\thanks{E-mail: ferrara@sissa.it} and Massimo Ricotti\thanks{E-mail: ricotti@astro.umd.edu}\\
SISSA/ISAS, via Beirut 2-4, 34014 Trieste, Italy \\
Dept. of Astronomy, University of Maryland, College Park, MD 20742-2421, USA}

\maketitle

\date{\today}

\begin{abstract}   
A model for gas outflows is proposed which simultaneously explains the correlations between the (i) equivalent width of 
low ionization and Ly$\alpha$ lines, (ii) outflow velocity, and (iii) star formation rate observed in Lyman Break Galaxies 
(LBGs). Our interpretation implies that LBGs host short-lived ($30 \pm 5$ Myr) starburst episodes observed at different 
evolutionary phases. Initially, the starburst powers a hot wind bound by a denser cold shell, which after $\approx 5$ Myr  
becomes dynamically unstable and fragments; afterwards the fragment evolution is approximately ballistic while the hot 
bubble continues to expand. As the fragments are gravitationally decelerated, their screening ability of the starlight 
decreases as the UV starburst luminosity progressively dims. LBG observations sample all these evolutionary phases.  
Finally, the fragments fall back onto the galaxy after $\approx 60$ Myr. This phase cannot be easily probed as it occurs 
when the starburst UV luminosity has already largely faded; however, galaxies dimmer in the UV than LBGs 
should show infalling gas.
\end{abstract}

\begin{keywords}
cosmology: theory -- galaxies: formation
\end{keywords}

\section{Introduction}
Recent observations (Adelberger \etal 2003, Shapley \etal 2003 (S03), Steidel \etal 2003, Adelberger \etal 2005) based on a combination of quasar absorption-line and faint-galaxy techniques applied to the same cosmic volumes have revealed that UV-selected, star forming galaxies at redshift $z\approx 3$ (in brief,  Lyman Break Galaxies, LBGs) show 
spectral absorption lines due to heavy elements. These absorption lines, which can be very strong ($N_\nCIV \gg 10^{14}$~cm$^{-2}$),  are often called ``interstellar" because they resemble the absorption lines produced by interstellar material in local galaxies. Nevertheless, these lines appear (without exception) to be  {\it blueshifted} with respect to the galaxy systemic velocity deduced from nebular lines. The typical velocity difference is $\Delta v \approx 250$~km~s$^{-1}$, but a non-negligible fraction of LBGs show differences in excess of 300 ~km~s$^{-1}$ (Pettini \etal 2002). These absorbing systems are often interpreted as arising in cool and/or neutral structures (maybe a shell) part of an outflow powered by supernovae associated with the detected galaxy star formation activity.  An additional element in support of this hypothesis comes from the detection of {\it redshifted} Ly$\alpha$ photons backscattered from gas behind the stars. Quite interestingly, this emission could trace the receding part of the outflowing shell.  
Finally, strong Ly$\alpha$ absorption is produced by the intergalactic gas within $1 h^{-1}$ comoving Mpc of most LBGs, but for about 1/3 the absorption is weak or absent. The transparency of the medium around some LBGs could arise from the gas collisional ionization/heating by the wind (Croft \etal 2002, Adelberger \etal 2003,  Kollmeier \etal 2003, Bruscoli \etal 2003, Bertone 2005, Kollmeier \etal 2006), from a galaxy proximity (photoionization) effect, as pointed out by Maselli \etal (2004), or simply reflect cosmic variance and artifacts of the data reduction (Desjacques \etal 2004, Desjacques, Haehnelt \& Nusser 2005).

In general,  one would expect to observe also infalling gas in the surroundings of LBGs arising either from gas accretion along the cosmic web filaments or because at least part of the outflowing gas could rain back onto the galaxy if its velocity does not exceed the escape speed of the system (a process known as galactic fountain). In fact, the infall ram pressure might be of primary importance in confining and reverting  the wind expansion (Fujita \etal 2004). This situation is clearly appreciated from the inspection of the results of recent, high resolution simulation of LBG outflows (Ferrara, Scannapieco \& Bergeron; see Fig. 2 of that paper). However, observations so far have not  found any experimental evidence of such physically-based expectation. 
     
The central aim of this paper is to understand why outflows are routinely observed in LBGs whereas gas infall seem to escape detection.  This puzzle is possibly related to the following scenario. During the first phase of a starburst the galactic wind produces a cavity filled with hot gas, enclosed by a cold and dense shell. Afterwards the shell fragments and the velocity of the cold debris starts
decreasing while the hot gas escapes from the galaxy.  Whether or not they survive for some time, the UV light from the galaxy should become progressively less absorbed by the debris since their covering factor decreases with time as they travel to larger distances from the galaxy.  Observations may show a time evolution of the equivalent width (EW) of metal absorption lines and dust reddening produced by the decrease of the screening effect of the cold debris on the UV light from the galaxy. If the debris are not disrupted or ejected from the galaxy we should also observe the gas return (infall) process. The relevant question to pose then seems: Can we observe galaxies 
during all these different evolutionary stages?

A large statistical galactic sample (more than 800 galaxies at intermediate redshifts) that can be used to test this model has been compiled by S03. From the analysis of this data set a very interesting number of relationships emerge (some of which already noticed by S03 and others obtained by correlating their data 
in a different way), which are listed in the following:
\begin{itemize}
\item[(1)] The wind velocity, $v_w$, is proportional to the star formation rate (SFR): $v_w  \propto \dot M_\star$
\item[(2)] The Ly$\alpha$ emission line EW is  inversely proportional to the wind velocity: ${\rm EW}_{Ly\alpha}\propto 1/v_w$
\item[(3)] The EW of metal absorption lines is proportional to the wind velocity: ${\rm EW}_{met}\propto v_w$
\item[(4)] The reddening is proportional to the wind velocity: $E(B-V)\propto v_w$
\end{itemize}

Should we interpret these properties as produced by intrinsic variations of the physical properties ({\it eg.}, SFR, mass, size) of
LBGs?  Or do they rather reflect the different ages of a single (broadly speaking) galactic population with uniform properties? The relationships (3) and (4) naturally descend from the properties of a time evolving uniform population: as the cold debris move outward they are slowed down by gravity/energy losses, their covering factor decreases, and the EW and reddening also decrease. Property (2) can easily be understood in the same framework.  Contrary to the metal absorption lines, the EW of the Ly$\alpha$ emission line increases with decreasing wind velocity ({\it ie.}, increasing time from the starburst) since the scattering by the outflowing neutral gas is reduced and a larger fraction of the blue wing of the Ly$\alpha$ line is transmitted. 

\begin{figure}
\centerline{\psfig{figure=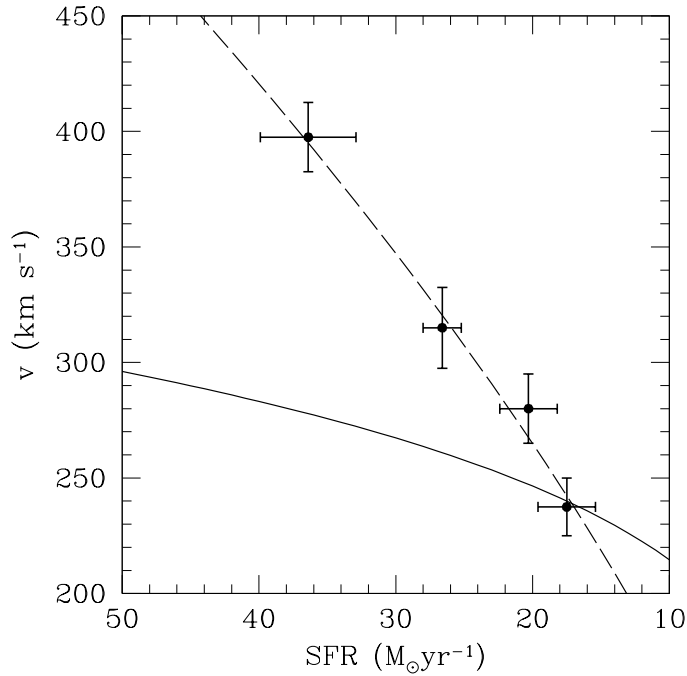,height=9cm}}
\caption{Wind velocity as a function of the LBG star formation rate. The points with errorbars show the data from S03; the dashed line is the best fit to the data with a power law $v_{w}   \propto \dot M_\star^{2/3}$.  A pure  luminosity effect would produce a much flatter power law:  $v_w  \propto \dot M_\star^{1/5}$ (solid line).}
\label{fig:v_sfr}
\end{figure}

Naively, relationship (1) can be interpreted as due to a variation of the energy input of the wind for galaxies with different star formation rate. However, a closer look at that relationship (\fig~\ref{fig:v_sfr}) shows that the dependence of $v_w$ on the SFR is not as the one expected under this assumption.  Assuming, for simplicity, a constant density profile for the gas in the galaxy core, the velocity is a very weak function of the SFR: $v \propto \dot M_\star^q$ with $q=1/5$ (shown by the solid line).  Even assuming a more general gas density profile of the galaxy core, $\rho_g \propto r^{-\beta}$, would produce $q =1/(5-\beta)$ that is too flat to fit the observations for any realistic choice of $\beta$.  A very steep slope $\beta=3.5$ (shown by the dashed line) fits the observations, but would produce unrealistically large outflow velocities incompatible with the data. Moreover, we will show later that the cold shell becomes unstable when the slope of the density profile becomes steeper than $\beta \simeq 2$: at that point  fragments decouple from the hot wind. Thus, the steepest physical dependence of $v_w$ on the SFR is $q \approx 1/3$  when $\beta=2$, while the best fit to the data requires $q \approx 2/3$.

Guided by the aforementioned arguments, in this paper we propose that the different observed wind velocities correspond to different starburst evolutionary times, and are not directly related to variations of their mean SFR or other internal properties. The proportionality of the wind velocity and SFR is incidental. It reflects the fact that both quantities decrease as the time elapsed from the starburst increases. 

We have organized the paper as follows. In \S~\ref{sec:sca} we describe our model analytically assuming a power law density profile
for the gas. In \S~\ref{sec:res} we assume a more realistic density profile of the gas and we compare the numerical results to the
observations. In \S~\ref{sec:dis} we discuss the implications of our model and provide an explanation for the lack of evidences for infalling gas in LBGs.

\section{Scaling arguments}\label{sec:sca}

To understand some basic features of LBG outflows we first resort to scaling arguments. We closely follow the method outlined in Ferrara \& Tolstoy (2000) and Madau, Ferrara \& Rees (2001) to which we refer the interested reader for the details. 
Assume that ambient gas pressure, gravity, and cooling can be neglected; moreover, we take a constant mechanical 
luminosity for the starburst, $L_w$ (the case $L_w \neq {\rm const}$ can be explored but there is no analytical 
solution). Then the outflowing (thin) shell radius evolution is described by:
\begin{equation} R_s(t) = \left({125\over 154 \pi}\right)^{1/5} \left({L_w    t^3\over\rho_g}\right)^{1/5}.
\label{rs1}
\end{equation}
We further make the hypothesis that the halo gas density profile can be approximated (at least piece-wise) by a power-law profile:
\begin{equation}
\rho_g(R) = \rho_{c}\left({R\over R_{c}}\right)^{-\beta} = \rho_{c} \left({0.22\over c}\right)^\beta x^{-\beta}  = \rho_{0} x^{-\beta}
\label{rho}, 
\end{equation}
with $x=R/R_{vir}$ and $\rho_{c}$ is the central density. As for a Navarro, Frenk \& White (1997, NFW) dark matter halo 
baryons have a core radius $R_c = (0.22/c) R_{vir}$, where $c$ is the concentration parameter, we have used such relation 
in eq. \ref{rho}. Note that $\rho_0\equiv \rho_c$ in the case of a central core, i.e. if $\beta=0$. 
By substituting $\rho_g$ into eq. \ref{rs1}  we find
\begin{equation}
x_s(t) = \left({125\over 154 \pi }\right)^{q} R_{vir}^{-5q} \left({L_w t^3\over \rho_{0}}\right)^{q},
\label{rs2}
\end{equation}
with $q=1/(5-\beta)$; we use the standard definition of $R_{vir}$
\begin{equation}
R_{vir}=24.7 \left ( \frac {M} {2\times 10^{11} h^{-1} {\rm M}_\odot} \right )^{1/3}
\left [ \frac{\Omega_m}{\Omega_m^z} \frac {\Delta_c}{18 \pi^2} \right ]^{-1/3}
\left ( \frac{1+z}{4} \right )^{-1} h^{-1} {\rm kpc}.
\end{equation}
We also take $\rho_{c} = \delta_{c}\langle \rho_g(z)\rangle
\approx \delta_{c} \Omega_b h^2 \rho_{crit}(1+z)^3$.  In the following, we adopt the value $\delta_c=2\times 10^4$, 
the appropriate overdensity of the core in a NFW profile. If we further define\footnote{We adopt the standard notation $Y_x=Y/10^x$}
\begin{equation}
t_w = \left[\left({125\over 154 \pi}\right){L_w \over \rho_{0}R^5_{vir}}\right]^{-{1 /3}}= 
2.5~{\rm Gyr}\left({0.22\over c}\right)^{\beta/3}\left({R_{vir}\over 24.7 ~{\rm kpc}}\right)^{5/3} \left({1+z \over 4}\right) L_{w,40}^{-1/3},
\label{tw}
\end{equation} 
eq.~(\ref{rs2}) can be casted in the very simple form:
\begin{equation}
x_s(t) =  \left({t\over t_w}\right)^{3q} = \tau^{3q}.
\label{rs3}
\end{equation}
The velocity\footnote{All lengths are normalized to $R_{vir}$, time to
  $t_w$, velocities to $R_{vir}/t_w$, masses to $t_w^3L/R_{vir}^2$} of
the shell is then given by
\begin{equation}
v_s(t) = {dx_s(t)\over dt} = 3q \tau^{3q-1} = 3q {x_s\over \tau}.
\end{equation}
The dense shell is pushing on the rarefied interstellar medium (ISM)
gas. The (external) shell interface becomes Rayleigh-Taylor (RT) unstable if the shell is
{\it accelerating}.  In this case ${\bf {a}}$ is parallel to ${\bf
v}$, and the vector goes from the heavy (shell) to the light (ISM)
fluid and the RT instability develops.  From MacLow \& McCray (1988)
simulations we note that the RT fingers are beyond the thin shell solution, i.e. the instability develops outside the bubble. The nonlinear evolution leads to the formation of clumps which detach from the shell; the hot gas wraps around them and escapes; at the same time, the ram pressure on the clumps decreases so they lag behind the shock.  Hence, the onset of the RT instability is fixed by the condition that the acceleration of the shell becomes positive. The shell acceleration is 
\begin{equation} 
g_s(\tau) = {dv_s(\tau)\over d\tau} = 3q(3q-1) \tau^{3q-2}. 
\end{equation} 
The RT condition is satisfied when $3q(3q-1) > 0$; as $q =1/(5-\beta) >0$ for any reasonable density profile slope $\beta$, this implies $\beta > 2$.  Hence the instability starts to develop at the time $\tau_i$ when the shell enters in the $\beta >2$ part of the gas distribution. Before that it evolves in a $\beta < 2$ density profile (e.g. the baryonic core in a NFW profile or the inner part of a disk).  The linear growth rate of the instability on spatial scale $\lambda=2\pi/k$ is given by
\begin{equation} 
\omega_{\rm RT}(\lambda,\tau) = 
\left[{{2\pi \vert g_s(\tau)\vert \over \lambda} {(\Delta - 1)\over
      (\Delta + 1)}}\right]^{1/2} \simeq \left[{2\pi \vert g_s(\tau)
    \vert \over \lambda }\right]^{1 \over 2},
\label{RTgrow1}
\end{equation}
the second equality descending by the fact that the density contrast between the shell and the ISM gas is $\Delta \gg 1$.  Note also that
the instability grows first on the shortest scales and it diverges as $\lambda \rightarrow 0$. At these small scales, the RT is stabilized
by viscosity and/or magnetic fields. As we are interested in the shell fragmentation, we estimate the growth time on a scale $\lambda$ such that $x_c = \lambda/ R_{vir}$, which by eq.~(\ref{rs3}) is equal to $\tau_i^{3q}$.  The maximum growth rate occurs when the acceleration is largest and scale is smallest, {\it i.e.}, at $\tau=\tau_i$. Then
\begin{equation} 
\tau_{RT}^{-1}(\tau_i) = \omega_{\rm RT}(\tau_i) = \left[{2\pi \vert g_s(\tau_i)\vert \over x_c}\right]^{1/2} 
                                = \sqrt{6\pi q\vert (3q-1)\vert} ~\tau_i^{-1} 
                                = {\sqrt{6\pi \vert (2-\beta)\vert }\over \vert 5-\beta \vert} ~\tau_i^{-1},
\label{RTgrow2}
\end{equation} 
where $0< \beta < 2$ is the value appropriate for the stable region. Given this range we find that the longest time corresponds to
$\beta=0$, for which $\tau_{RT}(\tau_i) = 0.8 \tau_i$.  So what is $\tau_i$? As already mentioned, the baryonic core in a NFW halo has size $x_c = 0.22/c$, hence for the $\beta=0$ ($q=1/5$) solution and for $c=5$ we find 
\begin{equation} 
\tau_i =x_c^{1/3q}
= x_c^{5/3} = (0.22/c)^{5/3} = 5.5\times 10^{-3}. 
\end{equation} 
At the time $\tau_f = \tau_i + \tau_{RT} = 1.8  \tau_i$, the instability becomes nonlinear and the shell fragments in dense clumps
the move almost ballistically in the gravitational field of the dark matter halo from there on.  Their precise mass/number is difficult to calculate, but to a first approximation $M_f \approx (\pi/6)\rho_{sh}\delta R_{sh}^3$, where $\delta R_{sh}$ is the shell thickness calculated at $\tau_f$; as now, we fix $M_f = \mu M_{sh}$ with $\mu < 1$. The mass of the shell is approximately equal to the swept-up mass in the halo.  To write the equation of motion for the fragments (note that this is {\it independent} of the mass of fragments) we first derive the nondimensional gravitational acceleration:
\begin{equation} 
g(x) = -{G M(R)\over R^2} {t_w^2 \over R_{vir}}= -
\left(t_w\over t_{ff}\right)^2 {F(cx)\over F(c)x^2}, 
\end{equation} 
where $t_{ff}\simeq \left[4\pi G \rho_{crit}\Omega_m(1+z)^3(200/3)\right]^{-1/2} \approx   0.32[(1+z)/4]^{-3/2}$~Gyr is proportional to the halo free fall time. We have assumed a NFW profile with 
\begin{equation} 
\rho(x)={\rho_{crit}\Omega_m(1+z)^3\,\delta\over cx (1+cx)^2},
\end{equation}
$\delta=(200/3)c^3/F(c)$, and
\begin{equation}
F(t)\equiv \ln(1+t)-{t\over 1+t}.
\end{equation}
It follows that fragment velocity is governed by 
\begin{equation}
{1\over 2} M_f {dv_f^2\over dx} = M_f g(x) = 
- M_f \left(t_w\over t_{ff}\right)^2 {F(cx)\over F(c)x^2}.
\label{eq:fra}
\end{equation}   
As $cx \ll 1$, $F(cx) \approx (cx)^2/(1+cx)$, then\footnote{In the  numerical results shown in \fig~\ref{fig:3} this  approximation has
been dropped.}
\begin{equation}
\int_{v_s(x_f)}^{v_f} dv^2 = - 2 \left(t_w\over t_{ff}\right)^2 
{c^2\over F(c)}\int_{x_f}^x  {dx \over (1+cx)} = -A \int_{x_f}^x  {dx \over (1+cx)},
\label{vel}
\end{equation} 
where $x_f=x_s(\tau_f)=(1.8\tau_i)^{3q}$, $v_s(x_f) = 3q\tau_f^{3q-1}=3q(1.8 \tau_i)^{3q-1}$ and the constant $A$ has value 
\begin{equation} 
A=2 \left(t_w\over  t_{ff}\right)^2 {c^2\over F(c)} = 2820.
\end{equation}
Note that $q \simeq 1/5$ ($\beta=0$) before $\tau_i$, but it increases to $q \simeq 1/3$ ($\beta=2$) after that, due to the steeper 
profile initiating the instability. We neglect that complication, by noting that the time-weighted mean is $q=0.3 \approx 1/3$ when computing  $x_f$ and $v_s(\tau_f)$. Thus, we obtain $x_f \approx \tau_f$ and $v_s(\tau_f) \approx 1$.
The resulting expression for the velocity is
\begin{equation} 
v_f^2(x) = v_s(x_f)^2 + A (x_f - x).  
\end{equation}
Substituting the expression for $v_s(x_f)$ above, and using $q=1/3$ we find \begin{equation} 
v_f(x)=\sqrt{1+A(x_f - x)}.  
\end{equation} 
The fragments are  decelerated and they stop at $x_0 = x_f + A^{-1}$ at a time $\tau_0$ obtained by solving 
\begin{equation}
  \int_{\tau_f}^\tau d\tau = \int_{x_f}^x {dx\over v} \rightarrow \tau
  = \tau_f + 2A^{-1}\left[1-\sqrt{1+A(x_f-x)}\right]; 
\end{equation}
evaluating this expression at $x=x_0$, yields $\tau_0 = \tau_f + 2 A^{-1}$. For $\tau > \tau_0$, the fragments start to infall onto the galaxy unless, $x_0 > 1$, in which case they are lost to the gravitational potential and ejected in the IGM. Using the value of $t_w$ from eq. \ref{tw} , with  $\beta=2$ and $c=5$ and the other fiducial values there,  one obtains $t_w=310 L_{w,40}^{-1/3}$~Myr and $A=52(t_w/t_{ff})^2=2820 L_{w,40}^{-2/3}$. Then we find
\begin{equation} 
\tau_0 = \tau_f + 2 A^{-1} = 1.8\tau_i + 7\times 10^{-4} L_{w,40}^{2/3} = 0.01 +
7\times 10^{-4} L_{w,40}^{2/3} 
\end{equation} 
\begin{equation} 
x_0 = x_f + A^{-1} = \tau_f + A^{-1}  = 0.01 + 7\times 10^{-4} L_{w,40}^{2/3}
\end{equation} 
We conclude that  mechanical luminosities $L_{w,40} > 1.5 \times 10^5$ are required to eject the fragments ($x_0>1$) from a typical LBG galaxy.  This conclusion is approximate as our formalism is valid in $cx \ll 1$ (see eq.  \ref{vel}) and it actually overestimates the velocity; nevertheless, it gives a first rough estimate of the escape of fragments.  
Are these mechanical luminosities consistent with the observed LBGs?  The mechanical luminosity (stellar winds and SNe) for a continuous star formation rate of $1 M_\odot$~yr$^{-1}$, with $M_{low} = 1 M_\odot, \alpha= 2.35, M_{up} = 100 M_\odot, Z=0.008$, is $\log L = 41.8$ (this slightly top-heavy IMF seems to be favored by the results obtained for high-redshift galaxies by Malhotra \& Rhoads 2003). Assuming that a fraction 
$\eta \approx 10\%$ (Veilleux\etal 2005) of this energy goes into kinetic energy, and the rest is radiated away, we find that $L_{w,40}=19 (\dot M_\star/30~{\rm M}_\odot~{\rm yr}^{-1})$. Thus an enormous SFR of about $230$ solar masses/yr would be needed (even more, given the overestimate of the velocity from eq.~[\ref{vel}]) to eject the fragment. The fact that cold, dense gas is very difficult to eject does not exclude that the hot gas can instead be expelled, as already noticed by Mac Low \& Ferrara (1999). 

Suppose that after shell fragmentation the clumps move radially and keep constant size (this is a reasonable assumption as long as they are embedded in and pressure-confined by the hot bubble interior gas flowing into the halo). Then their covering factor of the source (which is independent of their mass) decreases with time as 
\begin{equation}
{\cal C}(\tau) = {\left[x_f \over x_s(\tau)\right]}^2 = 
{\left[1.8  \tau_i\over x_s(\tau)\right]}^2 = 10^{-4} \tau^{-2},
\label{eq:c}
\end{equation}
that is, the covering factor has already dropped to 1\% when $x_s=10 x_f$, which occurs at $\tau = 10 x_f = 0.1$, or in dimensional units, after $250 L_{w,40}^{-1/3}$~Myr.

Finally, the hot shocked gas continues its expansion, leaving the clumps behind. Where does the shock stop? It stops when it reaches
pressure balance with the IGM surrounding the halo, i.e. when its velocity has dropped to the IGM effective sound speed, which is in the formula below is 
assumed to be $c_s=20$~km~s$^{-1}$. This happens
approximately when the internal pressure $L_wt_{burst}/R^3$ is equal to $\langle \rho_g(z)\rangle c_s^2$ or, at $z=3$,
\begin{equation}
R_s^{max} = \left[{L_w t_{burst}/\langle \rho_g(z)\rangle  c_s^2}\right]^{1/3} 
\approx \eta^{1/3} \left({\dot M_{\star} \over {\rm M}_\odot {\rm yr}^{-1}}\right)^{1/3} \left({t_{burst}\over 
10^8{\rm yr}}\right)^{1/3} {\rm Mpc}.
\end{equation}
The previous argument neglects radiative losses, gravity and the possible infall ram-pressure; hence, such estimate represents 
an upper limit to $R_s$ (physical units).

\section{Full Model}\label{sec:res}

The aim of this Section is to show quantitatively that the S03 data can be interpreted according to our model.  By
fitting the model to the data we attempt to infer some underlying properties of the LBG population. Hereafter, we adopt a more realistic description of the gas density profile with respect to the idealized case adopted in \S~\ref{sec:sca}. We also drop most of the simplifying approximations adopted previously. We solve numerically eq.~(\ref{rs1}) to calculate the evolution of the radius, velocity and acceleration of the shock front during the wind-driven phase. When the slope of the density profile becomes steeper than $\beta=2$ the cold shell becomes unstable to RT instability and fragments. The subsequent evolution of the cold fragments is calculated by solving eq.~(\ref{eq:fra}) and the time scale for shell fragmentation is calculated from eq.~(\ref{RTgrow1}). 
The quantitative results found in this Section are robust as they are insensitive to the poorly known properties of the gaseous disk. We will show that our model is independent of the particular choice of the gas density profile, at least for ``beta models'' or exponential density profiles. The results are also insensitive to the values of physical parameters such as the disk scale height, core density etc., provided that they obey a given relationship.

A gas in hydrostatic equilibrium in a NFW dark halo with virial radius $r_{\rm vir}$ and concentration $c$, has a density profile that is well fitted by a beta model (Makino, Sasaki \& Suto 1998)
\begin{equation} 
n_g={n_c \over [1+(R/R_c)^2]^{1.4}}, 
\end{equation}
where, for a gas at the virial temperature, the core radius is $R_c\simeq (0.22/c)R_{vir}$ and $n_c$ is the central gas
density. It is likely that, due to gas cooling, the temperature of the gas is lower than the virial temperature. In this case the core radius
will be smaller and the central density higher. Independently of the assumed value of $R_c$, if the gas mass is a cosmic fraction of the halo mass, $M_g=(\Omega_b/\Omega_{DM})M$, by imposing mass conservation and disk geometry we find
\begin{equation} 
n_c= 0.102~{\rm cm}^{-3}\left({R_c \over 1~{\rm kpc}}\right)^{-1}. 
\label{eq:loc}
\end{equation}
We have assumed the following fiducial values for LBGs: total mass
$M=2 \times 10^{11}$ M$_\odot$, concentration $c=5$ (\ie $R_c=1.32$ kpc). 

The data fitting is insensitive to the functional form of the density profile. We have found that adopting an exponential density profile
$n_{\rm g}=n_c \exp[-(R/R_c)^\alpha]$ (where now $R_c$ is the scale
height) the model fits the data equally well. The time scale for shell
fragmentation depends on the shape of the profile. The evolution of
the cold shell before fragmentation depends on the details of the
profile but the subsequent evolution of the cold fragments is
independent of the assumed profile. This happens because the fragmentation time-scale varies in such a way to keep the fragment velocity the same. 
We find that $\tau_{RT}/\tau_i=1.3$ for the ``beta
model'', $\tau_{RT}/\tau_i=0.93$ for the exponential profile with $\alpha=1$ and $\tau_{RT}/\tau_i=0.66$ for $\alpha=2$. During the acceleration and fragmentation phases, the velocity of the shell increases only by a factor $\Delta v/v_i$ of a few percent with
respect to its initial value $v_i$ at zero acceleration. This is consistent with eq.~(\ref{RTgrow1}) that can be recast in terms of the
variables of interest: $g_s \sim \Delta v/\tau_{RT}$ and $x_c \sim v_i \tau_i$. It follows that $\Delta v/v_i \sim (2\pi \tau_{RT}/\tau_i)^{-1} \ll 1$.

The velocity evolution of the cold fragments is degenerate for choices
of the mechanical luminosity of the starburst $L_w$ that obey $\eta
L_w = C n_c R_c^{3/2}$, where $C$ is a constant and $\eta$ is the fraction of mechanical luminosity converted into kinetic energy of the cold shell
(the rest being radiated away by the hot gas). Thus, further assuming
eq.~(\ref{eq:loc}), it follows that the velocity of the fragments
is proportional to $\eta L_{w} R_c^{-1/2}$. The cold shell driven by the starburst is stable for only a few Myrs before it
fragments and the clumps move ballistically. Thus the appropriate value of $L_w$ to use is that corresponding to this early phase, 
when the kinetic energy input produced by the starburst is dominated by winds from massive stars and is  $L_{w,40}=1-5 
(\dot M_\star/ M_\odot {\rm yr}^{-1})$, for stellar metallicities $Z/Z_\odot=0.4-1$ (Leitherer \etal 1999). The value of $\eta$, the fraction
of mechanical luminosity converted into kinetic energy of the supershell, is not well known from observations. We adopt the fiducial
value $\eta =10$\% (Vielleux \etal 2005). 

The LBG sample of S03 is  grouped in four bins according to their Ly$\alpha$ equivalent width. Each bin contains about the same number of galaxies (about 200). In our interpretation, different bins describe a statistically steady-state time sequence and correspond to different times from the beginning of the starburst with typical time scale $t_{burst}$.  Thus, if we neglect differential selection bias effects in the data, we expect that equally time-spaced bins contain an 
equal number of galaxies of age: $t_{i}=(i/4)t_{burst}$ for bins $i=1,2,3,4$.  The observed star formation rate of the galaxies in each of the four bins is well fitted
by an exponential function
\begin{equation} 
\dot M(t)=(29~{\rm M}_\odot~{\rm yr}^{-1}) \exp{\left(-{t \over t_{burst}}\right)}.
\label{eq:sfr}
\end{equation}
The numerical value of $t_{burst}$ will be derived later by fitting the observations of
the fragment velocities.  It also follows from eq.~(\ref{eq:sfr}) that the mean SFR integrated over the time scale of the burst is $\langle
\dot M_\star \rangle=0.632 \times \dot M(0)$ M$_\odot$ yr$^{-1}$ and the total stellar mass produced during the burst is $M_\star = \dot M(0)t_{burst}\approx 10^{9}~{\rm M}_\odot(t_{burst}/30 ~{\rm Myr})$.

\begin{figure}
\centerline{\psfig{figure=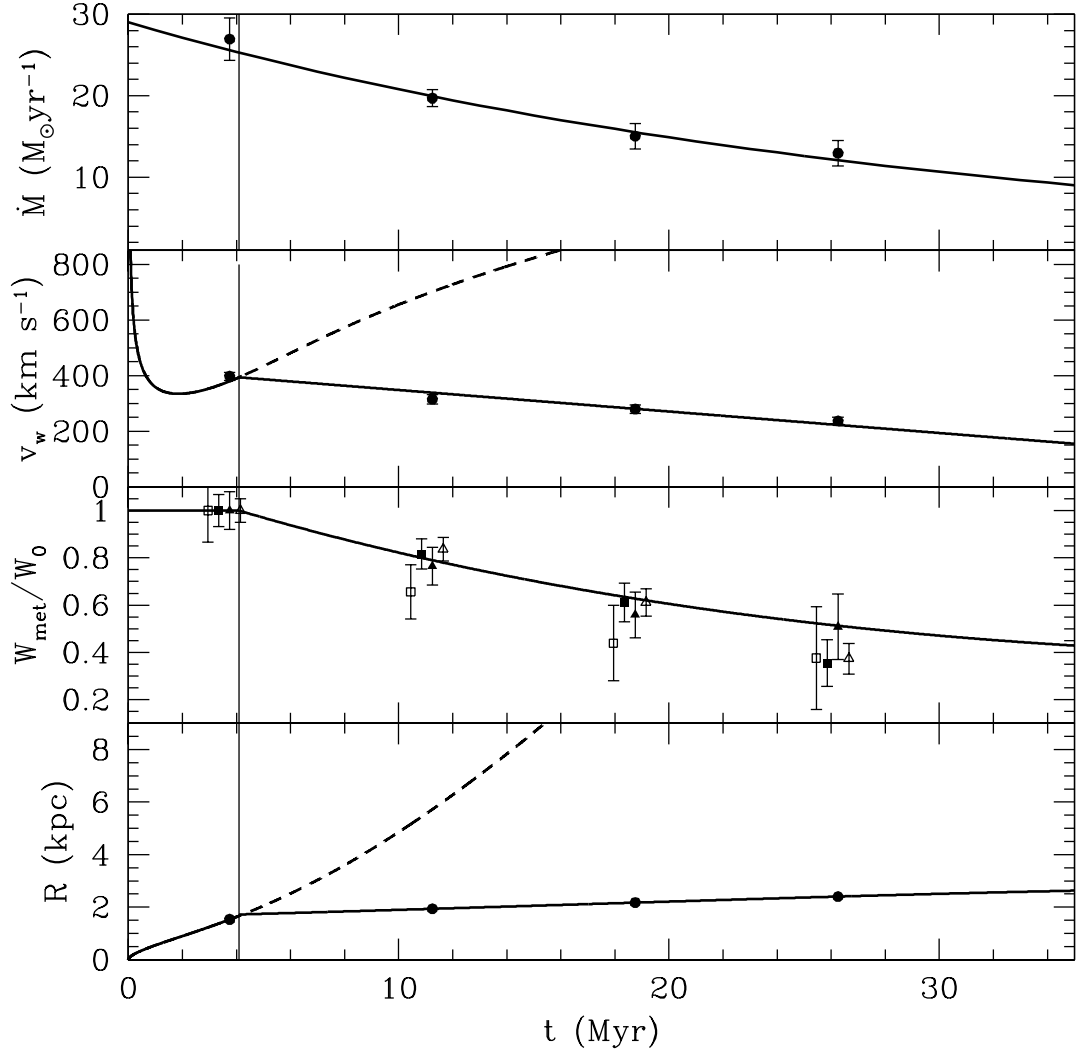,height=15cm}}
\caption{Outflow evolution as a function of time elapsed from the onset of LBG starburst. The symbols with errorbars show the S03 data (total of about 800 LBGs) grouped in four bins. The vertical lines show the time when shell fragments form and the hot gas freely escapes. From top to bottom the panels show the evolution of: (i) star formation rate,  (ii) hot gas (dashed) and fragment (solid) velocities, (iii) normalized equivalent width of low ionization metal lines (SiII, OI, CII and FeII), (iv) radius of the hot bubble (dashed) and position of the fragments (solid).}
\label{fig:3}
\end{figure}

In \fig~\ref{fig:3} we show the fit of our fiducial model (the ``beta model'' with $R_c=1$ kpc and $n_c=0.1$ cm$^{-3}$) to the data from
S03, shown by the symbols with errorbars. The panels from top to bottom show the time evolution of the star formation rate (fit by eq.~[\ref{eq:sfr}]), the velocity of the hot gas (dashed line) and cold shell/fragments (solid line), the equivalent width of low ionization metal absorption lines, and the distance from the galaxy of the hot (dashed line) and cold fragments (solid line). The symbols with errorbars in the third panel refer to the normalized EW of SiII (circles), OI (open circles), CII (squares) and FeII (open squares) calculated using eq.~(\ref{eq:c}).

Good fits to the observed velocity and EWs of the metal lines are obtained for values of the starburst time scale
\[t_{burst} \approx (30 \pm 5)~{\rm Myr}\] and for $L_{w,40}= 1 (\dot M_\star/M_\odot {\rm yr}^{-1})$ (appropriate for low stellar
metallicity) we find 
\[
\eta = 60 \% \left({n_c \over 0.1~{\rm
    cm}^{-3}}\right)\left({R_c \over 1~{\rm kpc}}\right)^{3/2}
    \approx 60 \% \left({R_c \over 1~{\rm kpc}}\right)^{1/2},
\]
where the second equality is derived by using eq.~(\ref{eq:loc}) that express mass conservation for the gas. As an example, 
if we instead assume an efficiency $\eta \sim 20$\% we find $R_c \approx 100$ pc, comparable to the disk scale height of the Milky Way. We also find that, 
independently of the assumed details of the gas density profile, the fragments 
start to fall back into the galaxy at $t_{fall} \approx 60$~Myr when they reach a distance from the disk of $2$ kpc. Although on larger scales the outflow approaches spherical geometry, the fragmentation point is located close enough to 
the galactic center that the effects of the disk density stratification are 
important.  

\section{Final considerations}\label{sec:dis}

We have already established that the cold clumps do not escape from LBGs. 
Therefore only two scenarios remain: either they disintegrate during the 
outflow or they eventually fall back onto the galaxy.  In the second case 
the reason why we do not see infalling clumps is simple: we do not observe 
this evolutionary phase because it happens too late when the UV emission 
from the galaxy has faded and the galaxy cannot be found using the Lyman break 
selection criteria (selection techniques based on infrared colors might instead
detect these galaxies long after the process has come to an end). More quantitatively, after about $50-100$ Myr the fragments will reach a maximum distance from the galaxy of about $2$ kpc and than start falling back.  According to our model the typical duration of the starburst phase is $t_{burst} \sim 30 \pm 5$ Myr.
Although ages since the most recent episode of star formation are found to be 
$\le 40$~Myr for 25\% of the LBG sample (Shapley \etal 2001), the derived  median value is 320 Myr. A possible explananation for this tension could be that the stellar light is affected by the integrated star formation history of the galaxy (which could be more complex than assumed in population synthesis models used for the interpretation of the data), while the kinematic properties of the wind are primarily affected by the very last burst episode. To clarify this point additional study is required.  

During this time scale the SFR declines as $\exp{[t/t_{burst}]}$ from 30 to 10 M$_\odot$~yr$^{-1}$.  During the starburst the fragments slow down from about $400$ km s$^{-1}$ to $100$ km s$^{-1}$. The UV burst is not long enough to allow the observation of galaxies showing the infall phase of the gas fragments. In our model we have assumed that the fragments are stable and survive for at least $\sim 30$ Myr, that is approximately the age of LBGs grouped in the last bin. In other words the galaxies in the last bin show the low ionization metal absorption lines that arise from dense gas in the fragments.

Let's now estimate the time scales for the destruction of the fragments.  The two main processes that may disintegrate the fragments
are shear instability (Kelvin-Helmholtz, hereafter KH) and conductive evaporation.  The time scale for the growth of the KH instability,
neglecting saturation effects, is approximately 
\begin{equation} 
t_{\rm KH}= 42~{\rm Myr}~\left({L_f \over 3~{\rm pc}}\right)\left({\Delta v \over 20~{\rm km s}^{-1}}\right)^{-1}\left({D \over 10^5}\right)^{1/2} \sim 50 - 100~{\rm Myr}, 
\label{eq:kh}
\end{equation} 
where $D=\rho_f/\rho_h$ is the fragment-hot gas density contrast, $L_f$ is the size of
the fragments and $\Delta v$ their velocity relative to the flow of hot gas. We have estimated the fragment size assuming that is
comparable to the thickness of the cold shell.  When $\Delta v/c_s>0.6$, the relevant case for our estimate, the instability growth
rate saturates and eq.~(\ref{eq:kh}) ceases to be valid (Vietri, Ferrara \& Miniati 1997).
In this case a good estimate of the minimum $t_{KH}$ is derived by plugging the value $\Delta v=0.2 c_s \sim 20$ km~
s$^{-1}$ in eq.~(\ref{eq:kh}) .  At least for the largest fragments we find that $t_{KH} \simeq t_{fall} > t_{burst}$. Magnetic
fields and cooling can increase the stability of the clumps, increasing our estimate of $t_{KH}$.  A dimensional estimate of
the evaporation time scale of the cold fragments embedded in the hot  wind as a result of conductive heating gives
\begin{equation} 
t_{\rm cond}= 2 k  K_e^{-1} n_f L_f^2=160~{\rm Myr} \left({N_f
      \over 3\times 10^{20} ~{\rm
      cm}^{-2}}\right)\left({L_f \over 3~{\rm pc}}\right)\left({T \over
      10^6~{\rm K}}\right)^{-5/2} \approx 150-300~{\rm Myr},
\end{equation} 
where $N_f=n_fL_f \approx \langle n_g\rangle x_f R_{vir}/3$ is the
mean column density of the fragments (assumed to be equal to the
column density of the cold shell).  We have assumed Spitzer conductivity
$K_e=1.5 \times 10^{-7} T^{5/2}$, neglecting saturation effects and
magnetic field (Spitzer 1953).

The survival time to KH instability and thermal evaporation are about
$100$ Myr, a factor of three larger than the life-time of the star
burst. The time scale for thermal evaporation is proportional to $L_f^2$ and for KH instability to $L_f$ . We expect that the larger clumps may survive long enough to start falling back after $t_{fall} \approx 60$ Myr but at this time the starburst has faded and the galaxy will not appear as a LBG with strong UV and Ly-alpha emission. It may be possible to test our model by selecting the faintest LBGs and searching for a subsample that has weak Ly$\alpha$ emission. This subsample may show evidence of either low velocity outflows or gas infall.

\vskip 1truecm
We would like to thank the referee, D. Weinberg, for insightful comments.


\begin{thebibliography}{}
\bibitem{} Adelberger, K. L., Steidel, C. C.,  Shapley, A. E., Pettini, M. 2003, ApJ, 584, 45
\bibitem{} Adelberger, K. L., Shapley, A. E., Steidel, C. C., Pettini, M., Erb, D. K.,  Reddy, N. A. 2005, ApJ, 629, 636
\bibitem{} Bertone, S, White, S. D. M. 2005, astro-ph/0511028
\bibitem{} Bruscoli et al. 2003, 2003, MNRAS, 343, 41
\bibitem{} Croft, R. A. C., Hernquist, L., Springel, V., Westover, M., White, M. 2002, ApJ, 580, 634
\bibitem{} Desjacques, V., Nusser, A., Haehnelt, M. G.,  Stoehr, F. 2004, MNRAS, 350, 879
\bibitem{} Desjacques, V., Haehnelt, M. G., Nusser, A. 2005,  astro-ph/0511025
\bibitem{} Ferrara, A., Scannapieco, E., Bergeron, J. 2005, ApJ, 634, 37 
\bibitem{} Ferrara, A.,  Tolstoy, E. 2000, MNRAS, 313, 291
\bibitem{} Fujita, A., Mac Low, M.-M., Ferrara, A.,  Meiksin, A. 2004, 2004, ApJ, 613, 159
\bibitem{} Kollmeier, J. A., Weinberg, D. H., Dav\'e, R., Katz, N. 2003, ApJ, 594, 75
\bibitem{} Kollmeier, J. A., Miralda-Escud\'e, J., Cen, R., Ostriker, J. P. 2006, ApJ, 638, 52
\bibitem[\protect\citeauthoryear{{Leitherer} et~al.}{{Leitherer}  et~al.}{1999}]{Leitherer:99}{Leitherer}, C., et~al. 1999, \apjs, 123, 3
\bibitem[\protect\citeauthoryear{{Mac Low} \& {McCray}}{{Mac Low} \&  {McCray}}{1988}]{MacLow:88}{Mac Low}, M.-M.,  \& {McCray}, R. 1988, \apj, 324, 776
\bibitem{} Mac Low, M.-M.,  Ferrara, A. 1999, ApJ, 513, 142
\bibitem{} Madau, P., Ferrara, A., Rees, M. J. 2001, ApJ, 555, 92
\bibitem[\protect\citeauthoryear{{Makino}, {Sasaki}, \& {Suto}}{{Makino}  et~al.}{1998}]{Makino:98}{Makino}, N., {Sasaki}, S.,  \& {Suto}, Y. 1998, \apj, 497, 555
\bibitem{} Malhotra, S., \& Rhoads, J. E. 2002, ApJ, 565, L71 
\bibitem{} Maselli, A., Ferrara, A., Bruscoli, M., Marri, S., Schneider, R. 2004, MNRAS, 350, 21
\bibitem{} Navarro, J. F., Frenk, C. S., White, S. M. N. 1997, ApJ, 490, 493  
\bibitem{} Pettini, Max, Rix, S. A., Steidel, C. C., Adelberger, K. L., Hunt, M. P., Shapley, A. E. 2002, ApJ, 569, 742 
\bibitem{} Shapley, A. E., Steidel, C. C., Adelberger, K. L., Dickinson, M., Giavalisco, M. \& Pettini, M. 2001, ApJ, 568, 95 
\bibitem{} Shapley, A. E., Steidel, C. C., Pettini, M.,  Adelberger, K. L. 2003, ApJ, 588, 65 (S03)
\bibitem[\protect\citeauthoryear{{Spitzer} \& {H{\"a}rm}}{{Spitzer} \&  {H{\"a}rm}}{1953}]{Spitzer:53}{Spitzer}, L.,  \& {H{\"a}rm}, R. 1953, Physical Review, 89, 977
\bibitem{} Steidel, C. C., Adelberger, K. L., Shapley, A. E., Pettini, M., Dickinson, M.,  Giavalisco, M. 2003, ApJ, 592, 728
\bibitem[\protect\citeauthoryear{{Veilleux}, {Cecil}, \&   {Bland-Hawthorn}}{{Veilleux} et~al.}{2005}]{Veilleux:05} {Veilleux}, S., {Cecil}, G.,  \& {Bland-Hawthorn}, J. 2005, \araa, 43, 769
\bibitem[\protect\citeauthoryear{{Vietri}, {Ferrara}, \& {Miniati}}{{Vietri}  et~al.}{1997}]{Vietri:97}{Vietri}, M., {Ferrara}, A.,  \& {Miniati}, F. 1997, \apj, 483, 262


\end{thebibliography}
\end{document}